\documentclass[english]{article}
\usepackage[T1]{fontenc}
\usepackage[latin9]{inputenc}
\makeatletter

\providecommand{\tabularnewline}{\\}

\makeatother

\usepackage{babel}
\begin{document}

\title{Attack on a classical analogue of the Dunjko, Wallden, Kent and Andersson
quantum digital signature protocol}

\author{Derrick Newton}

\maketitle
Faculty of Engineering and Computing, Coventry University, Engineering
and Computing Building, Gulson Road, Coventry, West Midlands CV1 2JH,
UK
\begin{abstract}
A quantum digital signature (QDS) protocol is investigated in respect
of an attacker who can impersonate other communicating principals
in the style of Lowe's attack on the Needham-Schroeder public-key
authentication protocol. A man-in-the-middle attack is identified
in respect of a classical variant of the protocol and it is suggested
that a similar attack would be effective against the QDS protocol.
The attack has been confirmed through initial protocol modelling using
a automated theorem prover, ProVerif.
\end{abstract}

\section{Introduction}

Traditional public and private key cryptographic protocols are usually
associated with the provision of confidentiality of message communication
between principals (usually identified as Alice and Bob) in respect
of some eavesdropping third-party (usually identified as Eve). However,
the existence of confidential channels of communication between principals
does not of itself necessarily imply authenticity or integrity of
the transmitted message, nor does it ensure non-repudiation of the
message by the sender. Signature schemes are implemented to provide
authenticity, integrity and non-repudiation of messages using public
key cryptography. Additionally, Lamport one-time signature schemes
\cite{Lamport1979} have also gained traction in practical situations
where processor overhead is of prime concern.

Quantum digital signatures (QDS) are a current topic for research
given that practical (but expensive) point-to-point quantum key distribution
(QKD) is commercially available. Dunjko et al have proposed BB84-based
one-time QDS schemes which do not require quantum memory \cite{Dunjko2014,Dunjko2014a}.
These protocols provide a QDS scheme free of the requirement for quantum
memory and processing resources, contrasting earlier schemes of Gottesman
and Chuang \cite{Gottesman2001}.

In this paper we investigate the QDS protocol P2 of Dunjko et al by
consideration of a classical analogy of the quantum protocol. We identify
a possible man-in-the-middle attack on the classical digital signature
protocol inspired by Gavin Lowe's attack on the Needham-Schroeder
protocol \cite{Lowe1995} and we propose a security attack on the
QDS protocol by analogy. Further, we are currently modelling the classical
P2 protocol in the Applied Pi Calculus and analysing the protocol
using an automated theorem prover called ProVerif. We discuss some
initial findngs from reachability experiments in support of our attacker
model.

\section{Protocol P2}

Dunjko et al's Protocol P2 is simplified and stated in classical form
as follows \cite{Dunjko2014a}:

\paragraph{Key distribution stage}
\begin{enumerate}
\item For each possible future message $m_{f}=0,1$, Alice ($A$) generates
two different secret keys consisting of sequences of classical bits.
\item For each possible message $m_{f}=0,1$, Alice sends one secret key
to Bob ($B$) and the other to Charlie ($C$) via secure classical
channels.
\item For each signature element and for $m_{f}=0,1$, Bob (Charlie) randomly
chooses to either keep it or send it to Charlie (Bob) via a secure
classical channel. Essentially, Bob (Charlie) applies a bitstring
mask operation against the key bitstring: $mask\left(k{}_{m_{f}B},n_{m_{f}B}\right)$
(resp. $mask\left(k{}_{m_{f}C},n_{m_{f}C}\right)$), where $n_{m_{f}B}$
(resp. $n_{m_{f}C}$) is Bob's (Charlie's) chosen bitstring mask for
future message bit $m_{f}$.
\end{enumerate}

\paragraph*{Messaging stage}
\begin{enumerate}
\item To send a signed one-bit message $m$, Alice sends $\left(m,k_{mB},k_{mC}\right)$
to Bob, say, where $k_{mB},k_{mC}$ are the secret keys assigned to
Bob (resp. Charlie) corresponding to the message $m$.
\item Bob checks $\left(m,k_{mB},k_{mC}\right)$ against his key and the
partial key sent to him by Charlie and accepts if matched.
\item Bob now forwards message $\left(m,k_{mB},k_{mC}\right)$ to Charlie
and Charlie checks this against his key and the partial key sent by
Bob and accepts if matched.
\end{enumerate}
Communication between the three principals over secure channels within
the protocol is represented by table \ref{tab:Classical-P2-protocol}.

\begin{table}
\begin{tabular}{llllllll}
1. & $A$ & $\longrightarrow$ & $C$ & : & $k_{0C},k_{1C}$ &  & \tabularnewline
2. & $A$ & $\longrightarrow$ & $B$ & : & $k_{0B},k_{1B}$ &  & \tabularnewline
3. & $B$ & $\longrightarrow$ & $C$ & : & $mask\left(k{}_{0B},n_{0B}\right),mask\left(k{}_{1B},n_{1B}\right)$ &  & \tabularnewline
4. & $C$ & $\longrightarrow$ & $B$ & : & $mask\left(k_{0C},n_{0C}\right),mask\left(k_{1C},n_{1C}\right)$ &  & \tabularnewline
5. & $A$ & $\longrightarrow$ & $B$ & : & $m,k_{mB},k_{mC}$ &  & \tabularnewline
6. & $B$ & $\longrightarrow$ & $C$ & : & $m,k_{mB},k_{mC}$ &  & \tabularnewline
 &  &  &  &  &  &  & \tabularnewline
\end{tabular}

\caption{\label{tab:Classical-P2-protocol}Classical P2 protocol}
\end{table}

\section{An attack on the P2 protocol}

In order to model the principle components of the P2 protocol we use
the classical statement of the protocol where point-to-point quantum
channels are replaced by secure classical channels.

\subsection{Possible attack by $E$ against $B$}

We will now consider a possible man-in-the-middle attack by Eve on
this protocol. Eve $\left(E\right)$ is a member of the communicating
network. She is trusted sufficiently by the other principals so that
secure communication channels can be established between her and her
fellow principals. Eve takes control over Bob's incoming and outgoing
communications so that Alice and Charlie send messages to Eve in the
belief that they are talking to Bob. Bob, on the other hand, receives
messages from Eve believing these messages to be originating from
Alice or Charlie. In this position Eve can choose to flip signatures
and message bits so that Bob receives a different message from Alice
than was originally sent and yet Bob is able to verify the message
against its attached signature. From Charlie's perspective, however,
the message received agrees with that which was sent by Alice. The
details of this attack are set out below and the communications are
presented in table \ref{tab:Man-in-the-middle-attack-against}.

\paragraph{Key distribution stage}
\begin{enumerate}
\item For each possible future message $m_{f}=0,1$, Alice ($A$) generates
two different secret keys consisting of sequences of classical bits.
\item For each possible message $m_{f}=0,1$, Alice sends one secret key
to Eve $\left(E\right)$ pretending to be Bob ($B$) and the other
to Charlie ($C$) via secure classical channels.
\item For each possible message $m_{f}=0,1$, Eve $\left(E\right)$ swaps
the secret keys so that the key $k_{0B}$ is assigned to future message
bit $m_{f}=1$ and key $k_{1B}$ is assigned to future message bit
$m_{f}=0$ (now identified as $k'_{0B},k'_{1B}$).
\item Bob applies random bit selection to his keys, i.e. he assigns bitstring
masks to each key for future message $m_{f}$ and sends the partial
keys $kpart'_{0B},kpart'_{1B}$ to Eve thinking that he is commmunicating
with Charlie.
\item Eve has knowledge of the complete keys for Bob and the partial keys
so she is able to compute the effect of the bitstring masks on the
original keys. She sends these ``restored'' partial keys $kpart_{0B},kpart_{1B}$
to Charlie who believes that he is receiving them from Bob.
\item Charlie sends his masked partial keys $kpart_{0C},kpart_{1C}$ to
Eve in the belief that he is sending them to Bob.
\item Eve swaps the partial keys sent by Charlie so that the partial key
asigned to future message $m_{f}=0$ is now assigned to $m_{f}=1$
and vice versa. The swapped partial keys $kpart'_{0C},kpart'_{1C}$
are then sent to Bob.
\end{enumerate}

\paragraph{Messaging stage}
\begin{enumerate}
\item Alice sends her one bit signed message $\left(m,k_{mB},k_{mC}\right)$
to Eve thinking that she is communicating with Bob.
\item Eve flips the message: $m=0$ is replaced by $m'=1$ (or vice versa).
\item Bob matches the signature to the swapped key $k'_{mB}$ and to the
swapped partial key from Charlie $kpart'_{mC}$ and accepts if matched.
\item Bob sends the flipped message and signatures $\left(m',k_{mB},k_{mC}\right)$
to Eve thinking that he is communicating with Charlie.
\item Eve flips the message again so that $m'=1$is replaced by $m=0$ (or
vice versa).
\item Charlie confirms the (original) message signature against his own
signature and against the partial signature received from Eve (assumed
to be Bob) which is the corrected partial signature consistent with
Alice's original signature.
\end{enumerate}

\paragraph{Result}

Following this attack Charlie has received the correct message $m$
from Alice and has assured himself of its authenticity by verification
of the signature. Bob has received the flipped message $m'$ from
Alice and assured himself of its authenticity by verification of the
signature. Consequently, authenticity and integrity of the message
have not been provided by the signature protocol. A similar attack
can be devised against Charlie.

\begin{table}
\begin{tabular}{llllllll}
1. & $A$ & $\longrightarrow$ & $C$ & : & $k_{0C},k_{1C}$ &  & \tabularnewline
2. & $A$ & $\longrightarrow$ & $E\left(B\right)$ & : & $k_{0B},k_{1B}$ &  & \tabularnewline
3. & $E\left(A\right)$ & $\longrightarrow$ & $B$ & : & $k'_{0B},k'_{1B}$ & swap $k_{0B},k_{1B}$ & \tabularnewline
4. & $B$ & $\longrightarrow$ & $E\left(C\right)$ & : & $kpart'_{0B},kpart'_{1B}$ &  & \tabularnewline
\multicolumn{7}{l}{where $kpart'_{0B}=mask\left(k'_{0B},n_{0B}\right),kpart'_{1B}=mask\left(k'_{1B},n_{1B}\right)$} & \tabularnewline
5. & $E\left(B\right)$ & $\longrightarrow$ & $C$ & : & $kpart_{0B},kpart_{1B}$ & swap $k_{0B},k_{1B}$ & \tabularnewline
\multicolumn{7}{l}{where $kpart_{0B}=mask\left(k{}_{0B},n_{0B}\right),kpart_{1B}=mask\left(k{}_{1B},n_{1B}\right)$} & \tabularnewline
6. & $C$ & $\longrightarrow$ & $E\left(B\right)$ & : & $kpart_{0C},kpart_{1C}$ &  & \tabularnewline
\multicolumn{7}{l}{where $kpart_{0C}=mask\left(k{}_{0C},n_{0C}\right),kpart_{1C}=mask\left(k{}_{1C},n_{1C}\right)$} & \tabularnewline
7. & $E\left(C\right)$ & $\longrightarrow$ & $B$ & : & $kpart'_{0C},kpart'_{1C}$ & swap partial keys & \tabularnewline
\multicolumn{7}{l}{where $kpart'_{0C}=mask\left(k{}_{1C},n_{1C}\right),kpart'_{1C}=mask\left(k{}_{0C},n_{0C}\right)$} & \tabularnewline
8. & $A$ & $\longrightarrow$ & $E\left(B\right)$ & : & $m,k_{mB},k_{mC}$ &  & \tabularnewline
9. & $E\left(A\right)$ & $\longrightarrow$ & $B$ & : & $m',k_{mB},k_{mC}$ & swap $m,not\left(m\right)$ & \tabularnewline
10. & $B$ & $\longrightarrow$ & $E\left(C\right)$ & : & $m',k_{mB},k_{mC}$ &  & \tabularnewline
11. & $E\left(B\right)$ & $\longrightarrow$ & $C$ & : & $m,k_{mB},k_{mC}$ & swap $m,not\left(m\right)$ & \tabularnewline
 &  &  &  &  &  &  & \tabularnewline
 &  &  &  &  &  &  & \tabularnewline
\end{tabular}

\caption{\label{tab:Man-in-the-middle-attack-against}Man-in-the-middle attack
against $B$ in the P2 protocol}

\end{table}

\section{Formal modelling of the protocol and the attack}

Research into formal modelling of the protocol is ongoing. However,
the findings of this paper are supported by reachability experiments
performed over the classical P2 protocol.

The classical P2 protocol is encoded in the Applied Pi Calculus of
Abadi and Fournet \cite{Abadi2001}, as modified by Blanchet, Smyth
and others and implemented in ProVerif \cite{Blanchet2015}. ProVerif
is a powerful automated protocol verification tool for reachability,
secrecy, correspondence properties and observational equivalence.
Experiments are performed with the protocol model to validate the
attack outlined above. Full details of the formal modelling analysis
will follow in a subsequent paper. However, initial experiments support
the existence of the attack as outlined above.

\section{Application to QDS }

The attack outlined above applies to the classical cryptographic interpretation
of protocol P2. However, it can be observed that the attack involves
reassignment of keys to future messages and application of Bob's key
mask so we suggest that the attack could be transferred to the QDS
protocol P2 in which key distribution is established using quantum
channels. No direct observation and subsequent collapse of the key
distribution qubits by Eve is required. It remains an open research
question as to whether or not compromise to quantum P2 or indeed to
protocol P1 can be verified by formal modelling.

In this research we have not, as yet, considered modelling using process
algebras developed for quantum communications. We are mindful of the
research which has been carried out in respect of quantum process
algebras by Gay and Nagarajan with CQP \cite{Gay2005}, Feng et al
with qCCS \cite{Ying2009} but the translation of correspondence and
reachability assertions to quantum protocol modelling has not, to
our knowledge been established to date. Indeed there are fundamental
questions as to the application of event labelling and process dependencies
to quantum protocol models which require resolution before we can
devise correspondence and reachability tests within quantum models.

\section{Conclusions}

In this paper we have presented an attack on the classical implementation
of Dunjko et al's signature protocol P2. The attack allows an eavesdropper
to modify a signed message and and swap signatures within the protocol
so that a principal is able to authenticate the message and signature
even though the message has been altered. Additionally, we suggest
that this attack would extend to a quantum digital signature version
of the P2 protocol as the eavesdropper is not required to observe
the key messages on the quantum channel.

A detailed analysis of the classical protocol using an automated theorem
prover, ProVerif is ongoing. Extending the work further to formal
modelling and analysis using quantum process algebras or modifications
to classical process algebras to encapsulate quantum processes is
a further goal of this research.

\bibliographystyle{plain}
\bibliography{newton}

\end{document}